\documentclass[journal, onecolumn, 11pt]{IEEEtran}
\usepackage{cite}
\usepackage{bm}
\usepackage{amssymb}
\usepackage{mathtools}
\usepackage[most]{tcolorbox}
\usepackage{enumerate}
\usepackage{graphicx}
\usepackage{hyperref}

\title{Diversity Multiplexing Trade-off and Selection Gain in Media-Based Modulation}
\author{Ehsan Seifi$^\dagger$ and Amir K. Khandani$^\ddagger$ \\
{\small $^\ddagger$ E\&CE Dept. University of Waterloo, Waterloo, ON, Canada, $^\dagger$ \& Ciena Inc. Ottawa, ON, Canada}} 

\date{}

\begin{document}
\maketitle

\begin{abstract}
The idea of Media-based Modulation (MBM), introduced in \cite{c0}\!\cite{isit2014}, is to embed information in the variations of the transmission media (channel states). Using an RF closure with $w$ RF walls, MBM creates a set of $2^w$ select-able states for the end-to-end channel. Each state represents an index of an MBM constellation point. In each state, the wave (tone) emanating from the transmit antenna experiences many pseudo-random back-and-forth reflections within the RF closure. The RF signal, upon finally leaving the RF closure, further propagates in the rich scattering environment to reach the receiver.  This results in an independent complex channel gain to each receive antenna.  As a result, coordinates  of different MBM constellation points (vectors of channel gains formed over received antennas) will be independent of each other. This is unlike legacy transmission schemes where a fixed constellation structure used at the transmitter will be multiplied by a fixed, but random, channel gain. 
Due to this independence property,  MBM offers several advantages vs. legacy systems, including  ``additivity of information over multiple receive antennas (regardless of the number of transmit antennas)'', and ``inherent diversity over a static fading channel''.  This work studies the Diversity-Multiplexing Trade-off (DMT)~\cite{DMT} of an MBM constellation. Analytical expressions are provided that demonstrate the advantages of MBM vs. legacy systems. In particular, it is shown that a $1\times N_r$ SIMO-MBM constellation equipped with an MDS code  (even with a relatively small code length) significantly outperforms an $N_r\times N_r$ legacy MIMO.
\end{abstract}

\section{Introduction}
Reference~\cite{c0} shows that embedding part or all of the
information in the (intentional) variations of the
transmission media (channel states) can offer
significant performance gains vs. traditional Single-Input Single-Output (SISO), Single-Input Multiple-Output (SIMO)
and Multiple-Input Multiple-Output (MIMO) systems. This method, coined in \cite{c0} as Media-Based Modulation (MBM), is in contrast with traditional wireless
systems where data is embedded in the
variations of an RF source (prior to the transmit antenna) to propagate via fixed propagation paths (media) to the destination.  In particular, using capacity arguments, reference \cite{c0} shows that by using a single transmit antenna and a single or multiple receive antennas; MBM can significantly outperform SBM.
Following~\cite{c0}, reference \cite{isit2014}
proves that, a $1 \times K$ MBM over a
static multi-path channel asymptotically achieves the capacity
of $K$ (complex) AWGN channels, where for each unit of energy
over the single transmit antenna, the effective energy for each
of the $K$ AWGN channels is the statistical average of channel
fading. It is shown that significant gains can be realized even in a SISO-MBM setup. An example for the practical realization of the system using RF mirrors, accompanied with realistic RF and ray tracing simulations, are presented. 

MBM falls within the more general category of ``index modulation" \cite{index}, which includes a number of different approaches for modulating the RF carrier in the spatial domain. In particular, spatial modulation and its generalizations rely on multiple ON-OFF transmit antennas, wherein part of the data selects one of the transmit antennas (to be ON) and the rest of the data modulates the carrier to be transmitted through the selected  antenna. As a result, the number of data bits embedded in the channel state will be equal to $\log_2$ of the number of ON-OFF antennas.  The primary difference between MBM and spatial modulation is that MBM relies on the use of multiple ON-OFF RF mirrors (parasitic RF elements with two states) placed around a transmitting antenna. Due to the interaction among RF mirrors, each selection of ON-OFF pattern for the mirrors results in a new pattern for the overall transmit antenna structure. Consequently, the number of data bits embedded in the channel state will be equal to the number of ON-OFF mirrors, resulting in a linear scaling in rate vs. the logarithmic scaling in the case of spatial modulation. Due to space limitations and sheer volume of publications on the topic of spatial modulation, readers are referred to \cite{index}, and references therein, for more details. 

This article studies the advantages of MBM scheme with respect to the achievable {\em diversity} and {\em spatial multiplexing} gains. While un-coded MBM offers a good trade-off between {diversity} and {multiplexing}, using a maximum distance separable (MDS) code with minimum distance $D$ increases {diversity} by a factor of $D$ for a small reduction in {multiplexing}. Larger code lengths attain such {diversity} gain with a smaller reduction in {multiplexing}. Next, we review the problem formulation and the outline of the results given in the paper.

\subsection{Diversity and Multiplexing} A scheme is said to achieve {spatial multiplexing gain} $r$ and {diversity gain} $d$ if it supports the data rate
\begin{align}
\lim_{\mathsf{snr} \to \infty} \frac{R(\mathsf{snr})}{\log \mathsf{snr}} = r \> \text{(bit/s/Hz)}
\end{align}
with the average error probability 
\begin{align}
\lim_{\mathsf{snr} \to \infty} \frac{\log P_e(\mathsf{snr})}{\log \mathsf{snr}} = -d,
\end{align}
where $\mathsf{snr}$ is the average signal-to-noise ratio (SNR) at each receive antenna \cite{DMT}. 

After providing the mathematical model for MBM system in section \ref{sec::model}, we obtain the pair-wise error probability over the ensemble of MBM realizations in section \ref{sec:exact}. Subsequently, section \ref{sec:uncoded_tradeoff} shows that un-coded media-based modulation achieves {diversity gain} $d = N_r - r$, where $N_r$ is the number of receive antennas and $ r, \ 0 \leq r \leq N_r$ is the {spatial multiplexing gain}.
Furthermore, in section \ref{sec:coded_tradeoff} we establish that coded media-based scheme using an MDS code with maximum likelihood decoding, achieves {diversity gain} $d = DN_r-{r}/{\tau}$, where $D$ is the code minimum distance and $\tau$ is the MDS code dimensionless rate. More precisely, using an MDS code over $GF(q)$ with parameters $(N, K, D=N-K+1)$ the average word error probability is bounded by
\begin{align}
P_w \leq \frac{2^{N}}{\sqrt{2 \pi N_r}}\left(1+\frac{\mathsf{snr}}{2}\right)^{-(DN_r-{r}/{\tau})}.
\end{align}

\subsection{Selection Gain}
In MBM A coding gain may be achieved by means of a minimal feedback from the receiver. If the transmitter is provided with the knowledge of the index of RF mirror patterns that induce the lowest energy over the receive antennas, the communication may be enhanced by removing such subset with low projected energy. This will require using additional RF mirrors to maintain the same rate of transmission. In section \ref{sec:selection_gain}, we give analytical expressions for the potential energy gain as a function of additional RF mirrors.


\section{System model \label{sec::model}}

\begin{figure}[tbp]
\centerline{\includegraphics[scale = 0.45]{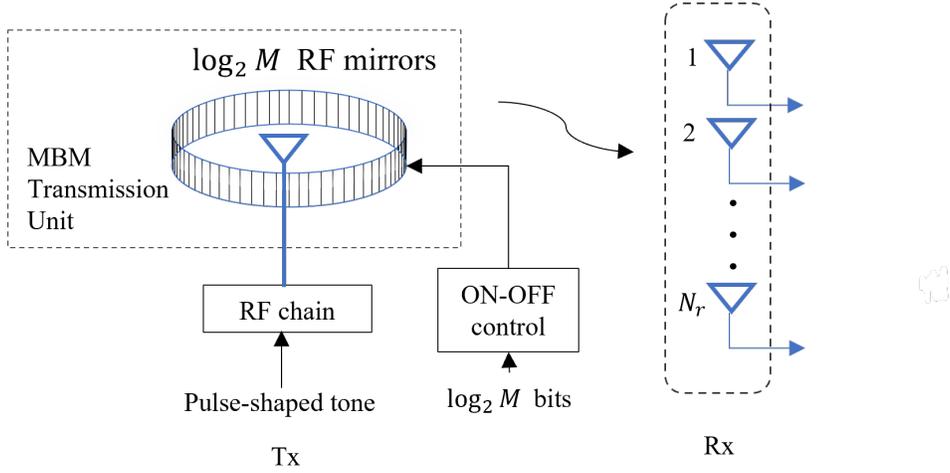}}
\caption{Media-based system.}
\label{fig:system_model}
\end{figure}

Fig. \ref{fig:system_model} shows a $1\times N_r = n/2$ SIMO-MBM system model with $N_r$ receive antennas and $n$ real dimensions at the receiver. There are $M$ messages and $\log_2 M$  RF mirrors. Each message index $m \in \{0, ..., M-1\}$ selects a unique ON-OFF pattern for the RF mirrors. Consequently, for each message a different channel gain $\mathbf{h}_m$ will be realized at the receiver. The projected components of channel gain along each receive dimension are $h_{m, d}, d = 1, ..., n$. 

We denote the set of all channel gains (due to all possible ON-OFF combinations of RF mirrors) by ${{\mathbf{h}}}^M := \{{\mathbf{h}}_0, {\mathbf{h}}_2, ..., {\mathbf{h}}_{M-1}\}$. Throughout the paper we will refer to  ${{\mathbf{h}}}^M$ as the set of MBM constellation points. In a static Rayleigh fading channel, the elements of MBM constellation set,  $\mathbf{h}_{m}$, are independent and identically distributed (i.i.d.) according to multivariate normal distribution $\mathcal{N}(0, \mathsf{snr}I_{n})$. Here, $\mathsf{snr}$ is the statistical average of signal energy at each receive dimension, i.e. $\mathbb{E}  h^2_{m, d} = \mathsf{snr}$. Even though the transmitter selects the message index $m$ and consequently $\mathbf{h}_m$, it is oblivious to its details. Receiver, however, knows $\mathbf{h}_m$. Receiver training is achieved by sending a set of pilot symbols. 

Assuming additive white Gaussian noise (AWGN), the noisy projected signals at the receiver $\mathbf{y}$ is written as
\begin{align}
\mathbf{y} = \mathbf{h}_m + \mathbf{z}.
\end{align}
AWGN noise vector $\mathbf{z}$ has i.i.d. components $z_d, d = 1, ..., n$. Without loss of generality, we assume noise variance per dimension is normalized to one: $\mathbb{E} z^2_d = 1$.


\section{Exact pairwise error probability \label{sec:exact}}
Using maximum likelihood decoding, given that message $m = 0$ is transmitted, (i.e. signal $\mathbf{h}_0$ is projected at the receiver), a decoding error occurs if
\begin{align}
\left\{\mathbf{h}_0 \rightarrow \mathbf{h}_i \> | \> \mathbf{h}_0 \right\}  =  \big \{ \lVert \mathbf{y} -\mathbf{h}_i \rVert^2 \leq \lVert \mathbf{y} -\mathbf{h}_0 \rVert^2 \big\},
\end{align}
for some $i \neq 0$.
Let ${\phi}(\mathbf{h}_0, \mathbf{h}_i)$ denote the pairwise error probability between $\mathbf{h}_0$ and $\mathbf{h}_i$ given $\mathbf{h}_0$ is sent. 
Since $\mathbf{h}_0$ and $\mathbf{h}_i$ are distributed normally, the squared Euclidean norm of their difference their follows a chi-squared distribution with $n$ degrees of freedom. Therefore, exact pairwise error probability averaged over ensemble of MBM realizations is obtained via
\begin{align}
\overline{{\phi}(\mathbf{h}_0, \mathbf{h}_i)} = \int_{x = 0}^{\infty} Q\left(\sqrt{\frac{\mathsf{snr} x}{2}}\right) f(x; n) \ \mathrm{d}x.
\label{integral_exact}
\end{align}
Here, the bar indicates averaging over ensemble, $f(x; n)$ is the probability density function for chi-squared distribution with $n$ degrees of freedom and $Q$ is the tail distribution function of the standard normal distribution. In \cite{789668} a closed form solution for this integral is provided according to
\begin{align}
\overline{{\phi}(\mathbf{h}_0, \mathbf{h}_i)} &= [P(c)]^{\frac{n}{2}} \sum_{k=0}^{\frac{n}{2}-1} \binom{\frac{n}{2}-1+k}{k} [1-P(c)]^k \\
P(c) &:= \frac{1}{2} \left( 1-\sqrt{\frac{c}{1+c}} \right),
\end{align}
where $c$ is defined as $c := {\mathsf{snr}}/2$. Same definition for $c$ is used throughout the article.
\section{Un-coded Diversity and Multiplexing \label{sec:uncoded_tradeoff}}
\begin{figure}[tbp]
\centerline{\includegraphics[scale  = 0.75 ]{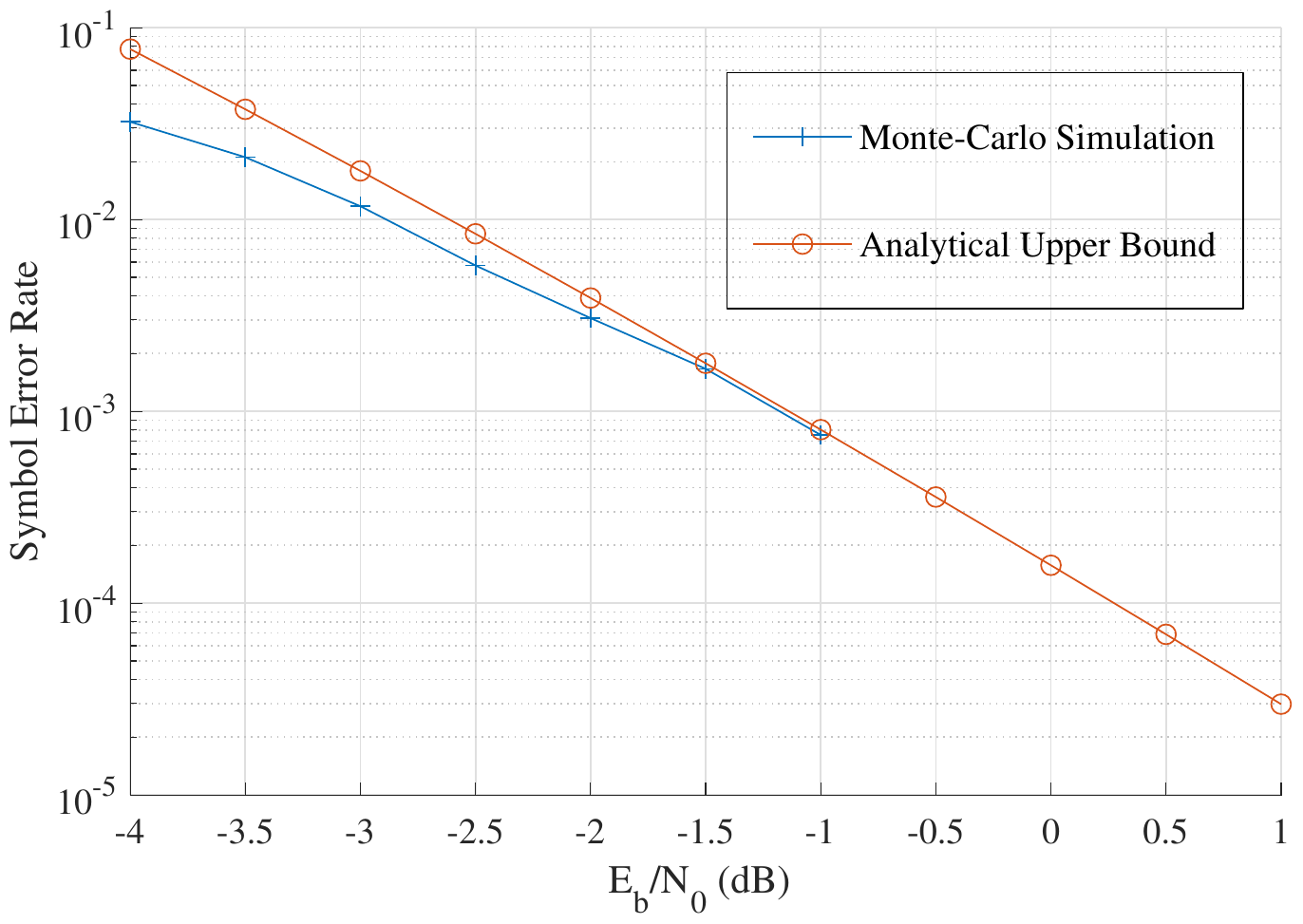}}
\caption{Monte-Carlo vs. bound \eqref{eq:union_bound} with $N_r = 8$ receive antennas and transmission rate $R = 16$ bits/sec/Hz. ${E_b}/{N_0} := {\mathsf{snr}}/{R}$.}
\label{fig:analytical_bound}
\end{figure}

We use the alternative form given in \cite{789668} to obtain a closed form bound on pair-wise error probability which does not require summation. The alternative form for $\overline{{\phi}(\mathbf{h}_0, \mathbf{h}_i)}$ is given by
\begin{align}
\frac{\sqrt{c/\pi} \Gamma(\frac{n+1}{2})}{2(1+c)^{\frac{n+1}{2}}\Gamma(\frac{n+2}{2})} \> _{2}F_1(1, \frac{n+1}{2}; \frac{n+2}{2}; \frac{1}{1+c}) \label{eq:closed_form},
\end{align}
where $_{2}F_1(.,.;.;)$ indicates  \textit{Gauss hyper-geometric function}. Applying the bound given in  \cite{gauss_ref} for {Gauss hyper-geometric function}, we get
\begin{align}
\overline{{\phi}(\mathbf{h}_0, \mathbf{h}_i)} \! \leq \frac{\Gamma(\frac{n+1}{2})}{ 2 \sqrt{\pi}\Gamma(\frac{n+2}{2})} (1+c)^{-\frac{n}{2}} \! \left(\!1+\! \frac{n+1}{(n+2)c} \right).
\label{eq:tight_upper_bound}
\end{align}
Using the union-bound and the fact that $\overline{{\phi}(\mathbf{h}_0, \mathbf{h}_i)}$ is independent of $\mathbf{h}_0$ and $\mathbf{h}_i$, the average probability of error over the ensemble of media-based constellation is bounded by
\begin{align}
P_e \leq \frac{ (M-1)\Gamma(\frac{n+1}{2})}{ 2 \sqrt{\pi}\Gamma(\frac{n+2}{2})} (1+c)^{-\frac{n}{2}} \left(1+ \frac{n+1}{(n+2)c} \right).
\label{eq:union_bound}
\end{align}
Fig. \ref{fig:analytical_bound} provides a comparison of \eqref{eq:union_bound} vs. simulation. Plugging $\log M = r \log \mathsf{snr}$, the average error probability at high SNR is closely approximated by
\begin{align}
P_e(\mathsf{snr}) \approx \frac{2^{\frac{n}{2}}\Gamma(\frac{n+1}{2})}{ \sqrt{\pi}\Gamma(\frac{n+2}{2})} (\mathsf{snr})^{-(\frac{n}{2} -r)}.
\end{align}
Recall that $N_r=2n$ is the number of receive antennas. Therefore, un-coded MBM achieves {diversity gain} $d = N_r - r$, while maintaining date rate $R = \log M = r \log \mathsf{snr}$.

\begin{figure}[b]
\centerline{\includegraphics[scale  = 0.75 ]{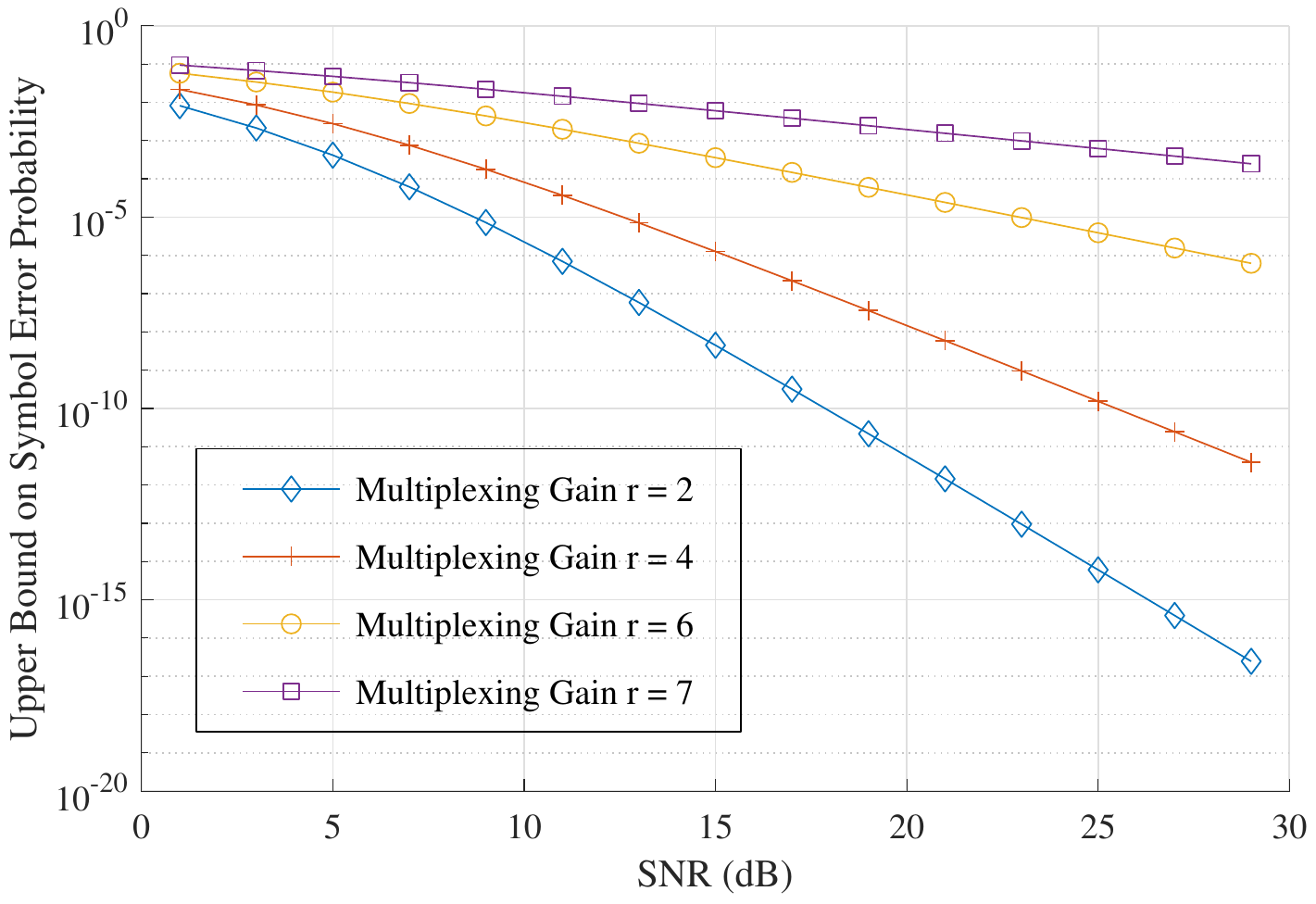}}
\caption{Error probability slope (i.e. diversity gain) for different rates (i.e. spatial multiplexing gains) for $N_r = 8$ receive antennas.}
\end{figure}

\section{Coded Diversity and Multiplexing \label{sec:coded_tradeoff}}
In this section, we follow the steps developed in \cite{1057352} and specialize them to media-based scheme to arrive at achievable diversity/multiplexing gains using MDS codes. For completeness, we will provide all the details.

Let $\Delta$ denote and MDS code over $GF(q)$, where $q$ is a prime power. Let $K$ and $N$, respectively, denote its dimension and block length. Then the minimum distance of $\Delta$ is given by $D=N-K+1$ and the dimensionless rate is $\tau = {K}/{N}$.
\subsection{Encoder Mapping}
The symbols of the code-words in $\Delta$ are mapped into sequences of MBM constellation points. To make the analysis simpler, we shall assume that the constellation set is partitioned into $N$ disjoint subsets. Each subset is assigned to one coordinate of code-words $\mathbf{u} = (u_1, u_2, ..., u_N) \in \Delta$.
This is to make sure, even identical symbols of $GF(q)$ in different coordinates of the code-word sequence are mapped to different points of the constellation set. Hence, we consider the constellation sets with cardinality greater than or equal to $Nq$. More specifically, the Galois field elements in the $i$th coordinate of a code-word are mapped to the $i$th subset of constellation points via mappings $g_i, i = 1, ..., N$. Then the projected code-words at the receiver has the form 
\begin{align}
\mathbf{h} = g(\mathbf{u}) := (g_1(u_1), ... , g_N(u_N)), 
\end{align}
for every $\mathbf{u} \in \Delta$. 

In a nutshell, the encoder operations are: 1) Map a message from the set ${1, 2, .., q^K}$ into an MDS code-word $\mathbf{u} \in \Delta$, 2) Map MDS code-word $\mathbf{u}$ into a sequence of MBM constellation points, $\mathbf{h} = g(\mathbf{u})$.

\subsection{Decoder Analysis}
In these analysis we consider word error probability of MDS codes over random ensemble of media-based constellation. Hence, for each fixed $\mathbf{u} \in \Delta$, the code-word $g(\mathbf{u})$ and the error probability ${\phi}(\mathbf{u})$ given $\mathbf{u}$ is sent, become random variables defined on ensemble, and we will use bar to indicate ensemble average.

The receiver is assumed to use maximum likelihood decoding. 
The noisy projected coded vector at the receiver is of the form
\begin{align}
\mathbf{y} = \mathbf{h} + \mathbf{z},
\end{align}
where
\begin{align}
\mathbf{h} = (\mathbf{h}_1, ..., \mathbf{h}_N), \ \mathbf{h}_i = (h_{i, 1}, ..., h_{i, n}) \>, \> 1 \leq i \leq N  \\
\mathbf{y} = (\mathbf{y}_1, ..., \mathbf{y}_N), \ \mathbf{y}_i = (y_{i, 1}, ..., y_{i, n}) \>, \> 1 \leq i \leq N 
\end{align}
In i.i.d Rayleigh fading, elements $h_{i, j}$ of the constellation points are sampled identically and independently from a normal distribution. 

Let $S(n) = \{1, 2, ..., N\}$. 
For $I \subseteq S(N)$, let $\Delta(I)$ indicate the subset of code-words that have non-zeros in the positions given by $I$ and zeros outside these positions. For $I \geq D$, and upper bound on $|\Delta(I)|$ is the number of code-words in $\Delta$ that have zeros outside the positions given by $I$. Since for every MDS code, any set of $K$ symbols form an information set, we have,
\begin{align}
|\Delta(I)| \leq q^{|I|-D+1}, \quad|I| \geq D.
\label{eq:bound_on_size}
\end{align}

Using union bound the error probability given that $\mathbf{u}$ is sent is bounded by
\begin{align}
\overline{\phi (\mathbf{u})} \leq \sum_{\mathbf{u}^{\prime} \in \Delta} \overline{\phi (\mathbf{u}, \mathbf{u}^{\prime})}.
\label{eq:union_bound_word}
\end{align}
Here, ${\phi}(\mathbf{u}, \mathbf{u}^{\prime}) $ is the pairwise error probability between code-words $\mathbf{u}$ and $\mathbf{u}^{\prime}$.

To utilize the MDS code property that every set of $K$ symbols form an information set, first partition the summation in \eqref{eq:union_bound_word} into summations over $\mathbf{u}^{\prime} \in \Delta(I) +\mathbf{u} $.
\begin{align}
\overline{\phi (\mathbf{u})} \leq \sum_{\substack{I \subseteq S(N)\\ D \leq |I| \leq N}} \sum_{\mathbf{u}^{\prime} \in \mathbf{u} + \Delta(I)} \overline{\phi(\mathbf{u}, \mathbf{u}^{\prime})}.
\end{align}
Denoting by $d := \lVert g(\mathbf{u})- g(\mathbf{u}^{\prime}) \rVert$, the  Euclidean distance between the two code-words,
we have
\begin{align}
\overline{\phi (\mathbf{u})} \leq \sum_{\substack{I \subseteq S(N)\\ D \leq  |I| \leq N}} \sum_{\mathbf{u}^{\prime} \in \mathbf{u} + \Delta(I)} \mathbb{E} \left[Q\left(\frac{\sqrt{\mathsf{snr}} d(\mathbf{u}, \mathbf{u}^{\prime}) }{2} \right) \right].
\end{align}
Because only the symbol positions with nonzero value in $ \mathbf{u} - \mathbf{u}^{\prime}$ contribute to $d^2(\mathbf{u}, \mathbf{u}^{\prime})$, it follows that for $\mathbf{u}^{\prime} \in \Delta +\mathbf{u} $,
 \begin{align}
d^2(\mathbf{u}, \mathbf{u}^{\prime}) = \lVert g_I(\mathbf{u})- g_I(\mathbf{u}^{\prime}) \rVert ^2.
\end{align}
Given that  $\mathbf{u}^{\prime} \in \Delta +\mathbf{u} $, random variable $d^2(\mathbf{u}, \mathbf{u}^{\prime})/2$ is sampled from chi-squared distribution with $n|I|$ degrees of freedom. This follows from the independence of symbols in an information set. Consequently, $\overline{\phi (\mathbf{u})}$ is bounded by
\begin{align}
\sum_{\substack{I \subseteq S(N)\\ D \leq |I| \leq N}} \sum_{\mathbf{u}^{\prime} \in \mathbf{u} + \Delta(I)} \int Q\left(\sqrt{\frac{\mathsf{snr}x}{2}}\right) f(x; n|I|) \ \mathrm{d}x \\
= \sum_{\substack{I \subseteq S(N)\\ D \leq  |I| \leq N}} |\Delta(I)| \int Q\left(\sqrt{\frac{\mathsf{snr}x}{2}}\right) f(x; n|I|) \ \mathrm{d}x.
\label{eq:phi_u_bound}
\end{align}
Since the right hand-side of \eqref{eq:phi_u_bound} is independent of $\mathbf{u}$, it is a bound on the media-based ensemble probability of error and is independent of probabilities with which the MDS code-words are used. Define $P_w := \overline{\phi (\mathbf{u})}$.
Plugging \eqref{eq:bound_on_size} in \eqref{eq:phi_u_bound} to upper bound $|\Delta(I)|$ , it follows
\begin{align}
P_w 
&\leq \!\sum_{\substack{I \subseteq S(N)\\ D \leq  |I| \leq N}}\! q^{|I|-D+1} \! \int \! Q\left(\sqrt{\frac{\mathsf{snr}x}{2}}\right)\! f(x; n|I|)  \mathrm{d}x \\
&= \sum_{i = D}^{N} \binom{N}{i} q^{i-D+1} \int Q \left(\sqrt{\frac{\mathsf{snr}x}{2}}\right) f(x; ni) \mathrm{d}x.
\end{align}
Let us use \eqref{eq:tight_upper_bound} to replace the integral. For large $c$, the term $( {n+1})/\big({(n+2)c\big)}\leq 1$. Therefore,
\begin{align}
P_w &\leq \sum_{i = D}^{N} \binom{N}{i} q^{i-D+1} \frac{\Gamma(\frac{ni+1}{2})}{ 2 \sqrt{\pi}\Gamma(\frac{ni+2}{2})} (1+c)^{-\frac{ni}{2}} \\
&\leq \frac{q^{-D+1}}{\sqrt{2\pi n}}  \sum_{i = D}^{N} \binom{N}{i} \left[q(1+c)^{-\frac{n}{2}}\right]^i, \label{eq:gamma_bound}
\end{align}
where in \eqref{eq:gamma_bound} we have used the inequality ${\Gamma(i+0.5)}/{\Gamma(i+1)} \leq {1}/{\sqrt{i}}$ and inequality $0 < {1}/{\sqrt{i}} \leq 1$ for $ i\geq 1$. If $q < (1+c)^{{n}/{2}}$, the largest geometric term in above sum is obtained for $i=D$ . Hence,
\begin{align}
P_w &\leq \frac{q^{-D+1}}{\sqrt{2\pi n}}  \sum_{i = D}^{N} \binom{N}{i} \left[q(1+c)^{-\frac{n}{2}}\right]^D  \\
&\leq \frac{q^{-D+1}}{\sqrt{2\pi n}}  \left[q(1+c)^{-\frac{n}{2}}\right]^D  \sum_{i = D}^{N} \binom{N}{i} \\
&\leq \frac{ 2^N }{\sqrt{2\pi n}} \ q \left[1+c\right]^{-\frac{nD}{2}}.
\label{eq:nchoosek_bound}
\end{align}
Step \eqref{eq:nchoosek_bound} uses the fact that $\sum_{i = D}^{N} \binom{N}{i} \leq 2^N$. 

\begin{figure}[tbp]
\centerline{\includegraphics[scale  = 0.75 ]{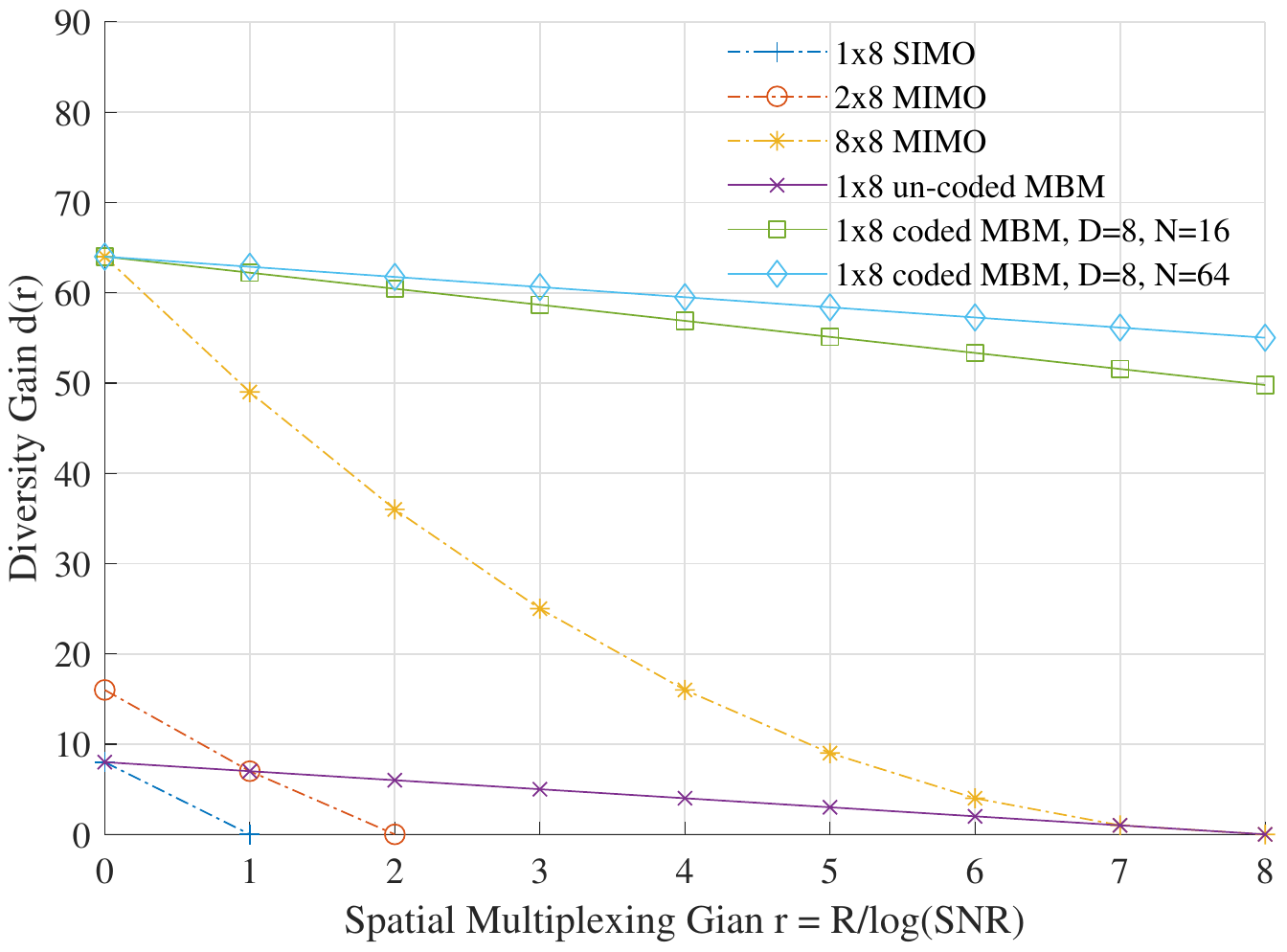}}
\caption{Diversity multiplexing trade-off, media-based vs. legacy MIMO.}
\label{fig:DMT}
\end{figure}

Let us choose the MDS code Galois field size according to $\log q = ({r}/{\tau}) \log \left(1+ {\mathsf{snr}}/{2} \right),  0 \leq {r}/{\tau} < N_r$. Such scheme achieves {spatial  multiplexing gain} $r$ while obtaining coded {diversity gain} $DN_r-{r}/{\tau}$. This statement is verified noting, 
\begin{align}
P_w \leq \frac{2^{N}}{\sqrt{2 \pi N_r}}\left(1+c\right)^{-(DN_r-{r}/{\tau})}.
\end{align}
Fig. \ref{fig:DMT} compares the achievable diversity and multiplexing gains in MBM vs. MIMO. 
A $1\times8$ MBM equipped with an MDS code of minimum distance $D=8$ (even with a relatively small code length) outperforms an $8\times8$ legacy MIMO.

\section{Selection gain due to removing low energy points \label{sec:selection_gain}}
\begin{figure}[btp]
\centerline{\includegraphics[scale  = 0.75 ]{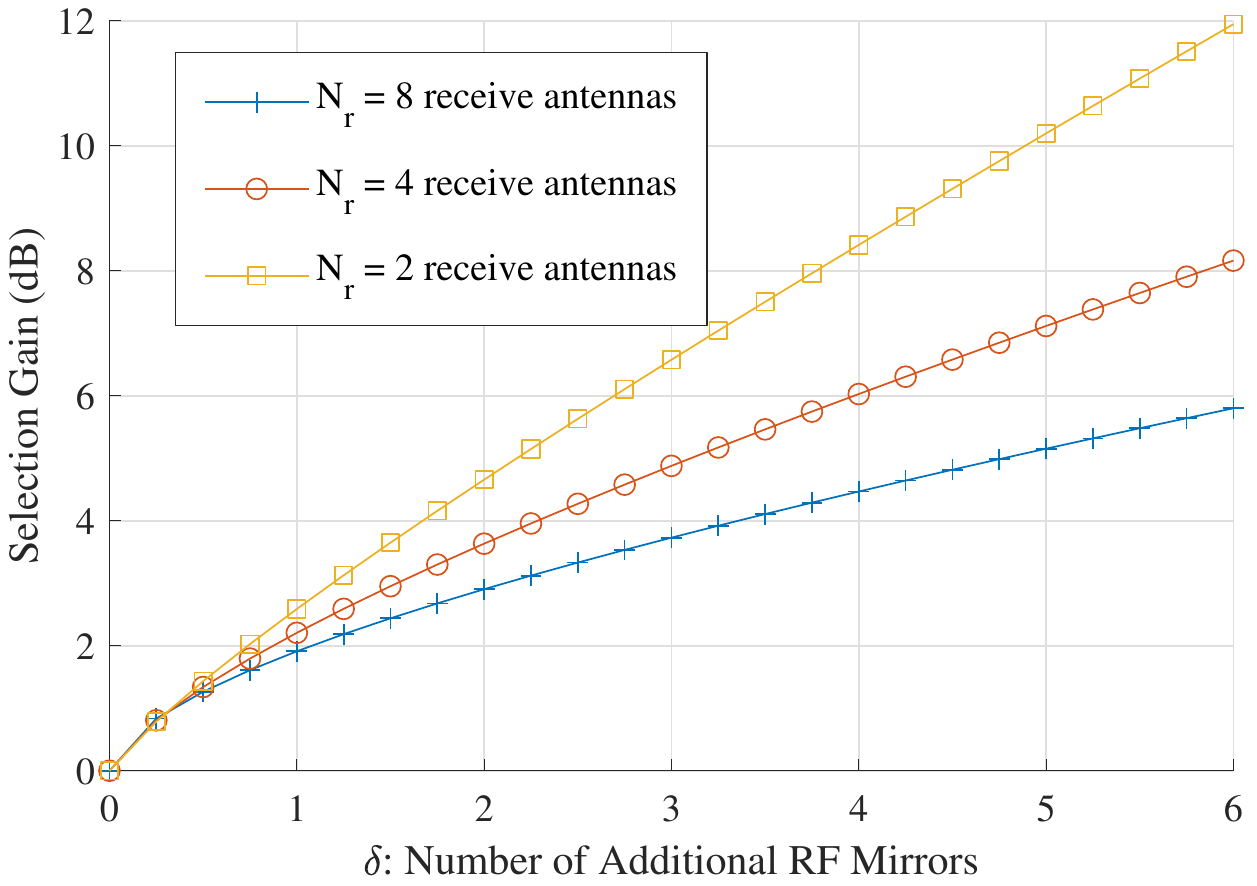}}
\caption {Selection gain (based on expressions in \eqref{eq:coding_gain}) for various number of receive antennas.}
\label{fig:selection_gain}
\end{figure}

\begin{figure}[btp]
\centerline{\includegraphics[scale  = 0.75]{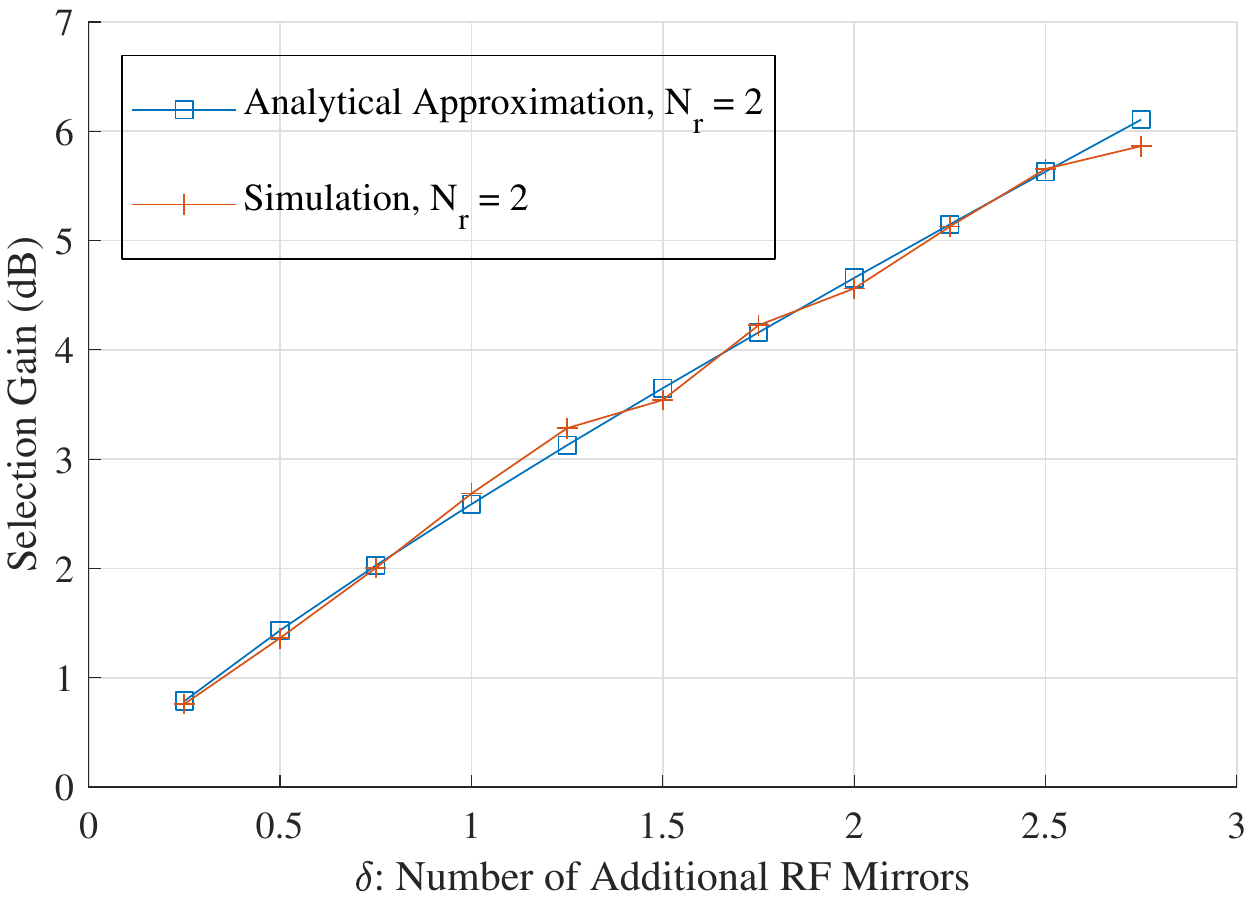}}
\caption{Comparison of analytical expression \eqref{eq:coding_gain} vs. simulated selection gain.}
\label{fig:selection_simulation}
\end{figure}

In MBM removing constellation points with the smallest projected energy over the receive dimensions provides coding gain. This is possible if the index of the points with the lowest energy is provided to the transmitter. Removing a subset of constellation points require using additional RF mirrors to maintain the same transmission rate (Each additional RF mirror increase the number of points by a factor of two). This section quantifies the potential energy gain as a function of average additional RF mirrors required to obtain the same rate.

For a given $E$, assume that the transmitter is provided with the index of the points in the constellation set with overall energy less than $nE$. Furthermore, the transmitter is instructed to remove those points from the constellation set. Assuming a maximum likelihood decoder, two consequences will follow:
\begin{enumerate}
\item \label{item1}  The required SNR to obtain the same average symbol error rate (averaged over all MBM ensembles in static Rayleigh fading channel) will approximately reduce by a factor of
\begin{align}
\label{eq:coding_gain}
\gamma_c = \bigg(\frac{\Gamma(\frac{n}{2}, \frac{n}{2}E)}{\Gamma(\frac{n}{2}, nE)}\bigg)^{\frac{2}{n}},
\end{align}
where $\Gamma(n, x) := \int_{x}^{\infty} t^{n-1}e^{-t} \mathrm{d} t$ indicates the upper incomplete gamma function.  
\item \label{item2} An average of $M \times \big(1-F(nE; n)\big)$ points will be removed from the constellation, where $F(x; n)$ is the cumulative distribution function for a chi-squared random variable with $n$ degrees of freedom. 
\end{enumerate}
Appendix \ref{appendix:app1} provides proof of items \ref{item1} and \ref{item2}.

SNR gain $\gamma_c$ can be thought of as a coding gain, and $\delta := \log_2 \big(1-F(nE; n)\big)$ as the average number of additional RF mirrors required to achieve an SNR gain specified in \eqref{eq:coding_gain} and maintain same rate of transmission. Fig. \ref{fig:selection_gain} illustrates savings in SNR (expressed in dB) as a function of number of additional RF mirrors, for various number of receive antennas. Fig. \ref{fig:selection_simulation} compares the accuracy of the analytic expression in \eqref{eq:coding_gain} versus Monte-Carlo simulation for $N_r = 2$ receive antennas.

\begin{appendices}
\section{Selection Gain Analysis}
\label{appendix:app1}

Fix an MBM constellation point $\mathbf{h}$ with energy $\lambda := \sum_{i=1}^{n} x_i^2 $. Given $\mathbf{h}$ is transmitted, the error probability averaged over competing symbol $\mathbf{h}^{\prime}$ is equal to
\begin{align}
P_e(\mathbf{h}) = \mathbb{E}_{\mathbf{h}^{\prime}} \left [ Q\left(\sqrt{\frac{\mathsf{snr} }{4}} \lVert \mathbf{h} -\mathbf{h}'\rVert \right) \right]. 
\label{eq:x_given}
\end{align}
Since $\mathbf{h}$ is given, random variable $\lVert \mathbf{h} -\mathbf{h}'\rVert^2$ follows a non-central chi-squared distribution with non-centrality parameter $\lambda$. Because $\mathbf{h}$ follows an isotropic Gaussian distribution, \eqref{eq:x_given} only depends on energy of the point, namely $\lambda$. In high SNR regime, we have
\begin{align}
P_e(\lambda) & := P_e(\mathbf{h}) \\
&= \int_{z=0}^{\infty} Q\left(\sqrt{\frac{\mathsf{snr} z}{4}}\right) f(z; n, \lambda) \ \mathrm{d}z \\
&=\sum_{i=0}^{\infty} \frac{({\lambda}/{2} )^i}{i!} e^{-\frac{ \lambda}{2}} \int_{z=0}^{\infty} Q\left(\sqrt{\frac{\mathsf{snr} z}{4}}\right) f(z; n+2i) \ \mathrm{d}z.
\end{align}
Replacing the integral with the closed form solution, gives
\begin{align}
P_e(\lambda)
&= e^{-\frac{ \lambda}{2}} \sum_{i=0}^{\infty} \frac{({\lambda}/{2})^i}{i!} \frac{\sqrt{{\mathsf{snr}}/(4\pi)}}{2(1+{\mathsf{snr}}/{4})^{\frac{n+2i+1}{2}}} \frac{\Gamma(\frac{n+2i+1}{2})}{\Gamma(\frac{n+2i+2}{2})} \> _{2}F_1(1, \frac{n+2i+1}{2}; \frac{n+2i+2}{2}; \frac{1}{1+{\mathsf{snr}}/{4}}) \\
&\approx \frac{1}{2 \sqrt{\pi}}e^{-\frac{ \lambda}{2}} (1+\frac{\mathsf{snr}}{4})^{-\frac{n}{2}}  \sum_{i=0}^{\infty} \frac{\left(\frac{ \lambda}{2(1+{\mathsf{snr}}/{4})}\right)^i}{i!}  \frac{\Gamma(\frac{n+2i+1}{2})}{ \Gamma(\frac{n+2i+2}{2})}.
\end{align}
The term in the summation is of the form of {\em confluent hypergeometric functions of the first kind}, $_1F_1(.; .; .)$. Accordingly,
\begin{align}
P_e(\lambda)
&= \frac{1}{\sqrt{2 \pi n}} e^{ -\frac{\lambda}{2}} (1+\frac{\mathsf{snr}}{4})^ {-\frac{n}{2}}   \ _1F_1(\frac{n+1}{2}; \frac{n+2}{2}; \frac{\lambda}{2(1+{\mathsf{snr}}/{4})}) \\
& \approx \frac{1}{\sqrt{2 \pi n}} e^{ -\frac{\lambda}{2}} (1+\frac{\mathsf{snr}}{4})^ {-\frac{n}{2}} (1+\frac{\lambda}{2(1+\mathsf{snr}/4)}) \  \\
&\approx \frac{1}{\sqrt{2 \pi n}} e^{ -\frac{\lambda}{2}} (1+\frac{\mathsf{snr}}{4})^{-\frac{n}{2}}.
\label{eq:given_energy}
\end{align}

To achieve a selection gain the transmitter is instructed to remove any constellation point with energy lower than $nE$. The average error probability given that low energy points are removed is given by
\begin{align}
P_e(E) &= \mathbb{E} \big[ P_e(\lambda) \ \big| \ \lambda \geq nE\big] \\
&= \frac{\Gamma(\frac{n}{2})}{\Gamma(\frac{n}{2}, \frac{nE}{2})} \int_{nE}^{\infty} P_e(\lambda) f(\lambda; n) \ \mathrm{d} \lambda.
\end{align}
Substituting $P_e(\lambda)$ with \eqref{eq:given_energy}, leads to
\begin{align}
P_e(E) 
&\approx \frac{\Gamma(\frac{n}{2})}{\Gamma(\frac{n}{2}, \frac{nE}{2})}  \frac{1}{\sqrt{2 \pi n}} \left(1+\frac{\mathsf{snr}}{4}\right)^{-\frac{n}{2}} \int_{nE}^{\infty} e^{ -\frac{\lambda}{2}} f(\lambda; n) \ \mathrm{d} \lambda \\
&= \frac{\Gamma(\frac{n}{2})}{\Gamma(\frac{n}{2}, \frac{nE}{2})}  \frac{1}{\sqrt{2 \pi n}} \left(1+\frac{\mathsf{snr}}{4}\right)^{-\frac{n}{2}} \int_{nE}^{\infty} e^{-\lambda} {\frac{\lambda^{\frac{n}{2}-1}}{2^{\frac{n}{2}}\Gamma(\frac{n}{2})}} \ \mathrm{d} \lambda  \\
&= {\frac{\Gamma(\frac{n}{2}, nE)}{2^{\frac{n}{2}}\Gamma(\frac{n}{2}, \frac{nE}{2})}} \frac{1}{\sqrt{2 \pi n}} \left(1+\frac{\mathsf{snr}}{4}\right)^{-\frac{n}{2}} \\
&\approx {\frac{\Gamma(\frac{n}{2}, nE)}{\Gamma(\frac{n}{2}, \frac{nE}{2})}} \frac{1}{\sqrt{2 \pi n}} \left(\frac{\mathsf{snr}}{2}\right)^{-\frac{n}{2}}.
\end{align}
Effectively, the required SNR to achieve the same symbol error probability is reduced by a factor given by, 
\begin{align}
\gamma_{c} \approx \left( \frac{\Gamma(\frac{n}{2}, \frac{nE}{2})}{\Gamma(\frac{n}{2}, nE)} \right) ^{\frac{2}{n}}.
\end{align}

The choice of $E$ determines the selection gain as well as the number of extra RF mirrors required to maintain the same rate of transmission. Let $M$ be the initial number of points, and $ M^{\prime}$ the remaining number of points after removing low energy ones. Since energy of an MBM constellation point follows a chi-square distribution, the average number of points remaining in the constellation set is given by the probability tail of chi-squared random variable with $n$ degrees of freedom.
\begin{align}
 M^{\prime} = M F(nE; n) 
\end{align}
Therefore, on average, $\delta = \log_2 \left( 1-F(nE; n) \right)$ additional RF mirrors is required to support the same transmission rate. 

\end{appendices}


\end{document}